\documentclass{llncs}
\usepackage{epsfig}
\usepackage{times}
\usepackage{theorem} 
\usepackage{graphs}
\usepackage{latexsym}
\usepackage{amstext}

\newlength{\picwidth}
\newenvironment{itemizeri}%
{\begin{list}{$\bullet$}{
   \labelwidth1ex 
   \labelsep1ex
   \rightmargin0ex
   \leftmargin2ex
   \listparindent0pt
   \itemsep0.2ex
   \topsep0.2ex
   \parsep0.1ex}}%
{\end{list}}

\theoremstyle{plain}    %
{\theorembodyfont{\rmfamily}
        \newtheorem{exxx}{Example}
}

\begin{document}
\bibliographystyle{abbrv}
\pagestyle{plain}

\title{Towards Rule-Based Visual Programming\\of Generic Visual Systems
        \thanks{This work has been partially 
                supported by  the ESPRIT Working Group
                \emph{Applications of Graph Transformation} (\textsc{Appligraph}).
               }
      }

\author{Berthold Hoffmann\inst{1} \and Mark Minas\inst{2}}
\institute{Fachbereich Mathematik/Informatik\\  
        Universit\"at Bremen\\  
        Postfach 33 04 40,   
        28334 Bremen, Germany\\
        hof@informatik.uni-bremen.de\\
        \and
        Lehrstuhl f\"ur Programmiersprachen\\ 
           Universit\"at Erlangen-N\"urnberg \\
           Martensstr. 3, 
           91058 Erlangen, Germany \\
        minas@informatik.uni-erlangen.de
}

\maketitle

\newcommand{\com}[1]{}
\newcommand{\change}[2]{\marginpar{#1} #2}
\newcommand{\ALT}[2]{#1}
\newcommand{\wzbw}{\hfill $\Box$}

\newcommand{\kw}[1]{\mathbf{#1}}
\newcommand{\fg}[1]{\textsf{#1}}

\newcommand{\tup}[1]{\langle #1 \rangle}
\renewcommand{\top}[1]{\widehat{#1}}
\newcommand{\cut}[1]{\underline{#1}}

\newcommand{\To}{\Rightarrow}
\newcommand{\notTo}{\not\Rightarrow}
\newcommand{\iso}{\cong}
\newcommand{\tuple}[1]{\langle #1 \rangle}
\newcommand{\G}{{\cal G}}
\newcommand{\E}{{\cal E}}
\renewcommand{\H}{{\cal H}}
\renewcommand{\L}{{\cal L}}
\newcommand{\N}{{\cal N}}
\renewcommand{\S}{{\cal S}}
\newcommand{\T}{{\cal T}}
\newcommand{\allows}{\vdash_{\S}}
\newcommand{\att}{\text{att}}
\newcommand{\lab}{\text{lab}}
\newcommand{\occ}{\text{occ}}
\newcommand{\glue}{\mathop{\!\not\!\circ\:}}
\newcommand{\guard}{\mathop{[\!]}}
\newcommand{\others}{\mathop{/\!\!/}}

  \graphnodesize{2} \graphnodecolour{1}         %
  \graphlinewidth{0.2} \grapharrowlength{1.5}   %
  \grapharrowwidth{0.8}  \grapharrowtype{1}     %
  \enlargeboxes{2} \autodistance{0.6}           %

\newcommand{\node}[3]%
  {\roundnode{#1}(#2,#3)}

\newcommand{\Node}[3]%
  {\roundnode{#1}(#2,#3)[\graphlinewidth{0.5}]}

\newcommand{\nnode}[5]%
  {\roundnode{#1}(#2,#3)\autonodetext{#1}[#4]{\small #5}}

\newcommand{\nnnode}[5]%
  {\roundnode{#1}(#2,#3)\autonodetext{#1}[n]{\small #4}
   \autonodetext{#1}[s]{\small #5}}

\newcommand{\point}[3]%
  {\roundnode{#1}(#2,#3)[\graphnodecolour{0}]}

\newcommand{\npoint}[5]%
  {\roundnode{#1}(#2,#3)[\graphnodecolour{0}]\autonodetext{#1}[#4]{\small #5}}

\newcommand{\nnpoint}[5]%
  {\roundnode{#1}(#2,#3)[\graphnodecolour{0}]\autonodetext{#1}[n]{\small #4}
   \autonodetext{#1}[s]{\small #5}}

\newcommand{\hiddennode}[3]%
  {\roundnode{#1}(#2,#3)[\graphnodesize{0}\opaquetextfalse]}

\newcommand{\smalledge}[4]%
  {\rectnode{#1}[4,4](#2,#3) \autonodetext{#1}{\footnotesize \fg{#4}}}

\newcommand{\Smalledge}[4]%
  {\rectnode{#1}[4,4](#2,#3)[\graphlinewidth{0.5}]
   \autonodetext{#1}{\footnotesize \fg{#4}}}

\newcommand{\var}[4]%
  {\rectnode{#1}[6,6](#2,#3) \autonodetext{#1}{\small $#4$}}

\newcommand{\typ}[4]%
  {\rectnode{#1}[6,6](#2,#3) \autonodetext{#1}{\footnotesize \fg{#4}}}

\newcommand{\framevar}[4]%
  {\rectnode{#1}[6,6](#2,#3) \autonodetext{#1}{$#4$}}

\newcommand{\Iframe}[6]%
  {\rectnode{#1}[#2,#3](#4,#5)[\graphnodecolour{0.7}] \autonodetext{#1}{#6}}

\newcommand{\Qframe}[6]%
  {\rectnode{#1}[#2,#3](#4,#5)[\graphnodecolour{0.85}]
        \autonodetext{#1}{\begin{graph}#6 \end{graph}}}

\newcommand{\button}[4]
  {\rectnode{#1}[14,5](#2,#3)[\graphnodecolour{1}\graphlinecolour{1}]
   \freetext(#2,#3){\begin{picture}(0,0)
                    \put(0,0){\oval(14,5)}\end{picture}}[\opaquetextfalse]
   \autonodetext{#1}{\makebox(0,0){\scriptsize \sf #4}}}

\newcommand{\buttonvar}[4]
  {\rectnode{#1}[7,5](#2,#3)[\graphnodecolour{1}\graphlinecolour{1}]
   \freetext(#2,#3){\begin{picture}(0,0)
                    \put(0,0){\oval(7,5)}\end{picture}}[\opaquetextfalse]
   \autonodetext{#1}{\makebox(0,0){\footnotesize $#4$}}}

\newcommand{\parbuttonvar}[4]
  {\rectnode{#1}[7,5](#2,#3)[\graphnodecolour{1}\graphlinecolour{0}]
   \freetext(#2,#3){\begin{picture}(0,0)
                    \put(0,0){\oval(6.8,4.8)}
                    \put(0,0){\makebox(0,0){\footnotesize $#4$}}
                    \end{picture}}[\opaquetextfalse]
   }

\newcommand{\parbutton}[4]
  {\rectnode{#1}[14,5](#2,#3)[\graphnodecolour{1}\graphlinecolour{0}]
   \freetext(#2,#3){\begin{picture}(0,0)
                    \put(0,0){\oval(13.9,4.8)}
                    \put(0,0){\makebox(0,0){\scriptsize \sf #4}}
                    \end{picture}}[\opaquetextfalse]
   }

\newcommand{\Edge}[2]%
  {\edge{#1}{#2}[\graphlinewidth{0.5}]}

\newcommand{\Bow}[3]%
  {\bow{#1}{#2}{#3}[\graphlinewidth{0.5}]}

\newcommand{\metaedge}[2]
  {\edge{#1}{#2}[\graphlinewidth{0.6}]\edge{#1}{#2}[\graphlinecolour{1}]}

\newcommand{\metabow}[3]
  {\bow{#1}{#2}{#3}[\graphlinewidth{0.6}]\bow{#1}{#2}{#3}[\graphlinecolour{1}]}

\newcommand{\ItemI}{%
\begin{graph}(7,7)(-3.5,-3.5)
  \node{0}{2.5}{0} \edge{0}{1} \edge{0}{2} \edge{0}{3}
  \node{1}{0}{-2} \edge{1}{3}
  \node{2}{0}{+2} \edge{2}{3}
  \point{3}{-2.5}{0}
\end{graph}%
}

\newcommand{\ItemII}{%
\begin{graph}(7,7)(-3.5,-3.5)
  \point{0}{2.5}{0} \edge{0}{1} \bow{0}{1}{-0.3} \bow{0}{1}{+0.3}
  \point{1}{-2.5}{0}
\end{graph}%
}

\newcommand{\ItemIII}{%
\begin{graph}(7,7)(-3.5,-3.5)
  \node{0}{2}{0} \edge{0}{1} \edge{0}{2} 
  \node{1}{-1.5}{-2} \edge{1}{2}
  \node{2}{-1.5}{+2} 
\end{graph}%
}

\newcommand{\Garrow}[3]%
{\begin{picture}(#1,#2) \put(0,0){\makebox(#1,#2){$#3$}}\end{picture}}

\newcommand{\Graephchen}[3]%
{\begin{picture}(#1,#2) \put(0,0){\makebox(#1,#2){#3}}\end{picture}}

\newcommand{\BILD}[3] 
{\begin{picture}(#1,#2) \put(0,0){\makebox(#1,#2){#3}}\end{picture}}

\newcommand{\DEF}[1] %
{\BILD{7}{#1}{::=}}

\newcommand{\OR}[1] %
{\BILD{3}{#1}{$\mid$}}

\newcommand{\IF}[1] %
{\BILD{3}{#1}{$[\!]$}}

\newcommand{\TO}[1] %
{\BILD{5}{#1}{$\to$}}

\newcommand{\TRANS}[1] %
{\BILD{7}{#1}{$\To$}}

\newcommand{\OUT}[2] %
{\BILD{10}{#2}{$\others \; #1$}}

\newcommand{\RULE}[3] %
  {\centering\begin{tabular}{ccc}
  #1 & \TO{#3} & #2 
  \end{tabular}}

\newcommand{\CONDITIONAL}[4] %
  {\centering\begin{tabular}{ccccc}
  #1 & \IF{#4} & #2 & \TO{#4} & #3 
  \end{tabular}}

\newcommand{\STEP}[3] %
  {\centering\begin{tabular}{ccc}
  #1 & \TRANS{#3} & #2 
  \end{tabular}}

\newcommand{\STEPS}[4] %
  {\centering\begin{tabular}{ccccc}
  #1 & \TRANS{#4} & #2 & \TRANS{#4} & #3 
  \end{tabular}}

\newcommand{\THREESTEPS}[5] %
  {\centering\begin{tabular}{ccccccc}
  #1 & \TRANS{#5} & #2 & \TRANS{#5} & #3 & \TRANS{#5} & #4
  \end{tabular}}

\newcommand{\PROC}[5]%
  {\begin{tabular}{@{}cccc@{}}
  #2 & \TO{#5} & #3 & \OUT{#4}{#5}
  \end{tabular}}

\newcommand{\CONPROC}[6]%
  {\begin{tabular}{@{}cccccc}
   #2 & \IF{#6} & #3 & \TO{#6} & #4 & \OUT{#5}{#6}
  \end{tabular}}

\newcommand{\TYPE}[4] %
  {\centering\begin{tabular}{@{}c@{}c@{}c@{}c@{}c@{}}
  #1 & \DEF{#4} & #2 & \OR{#4} & #3 
  \end{tabular}}

\newcommand{\SIGN}[3] %
  {\centering\begin{tabular}{ccc}
  #1 & \DEF{#3} & #2  
  \end{tabular}}

\newcommand{\TYPES}[6] %
  {\centering\begin{tabular}{@{}c@{}c@{}c@{}c@{}c@{}c@{}c@{}c@{}c@{}}
  #1 & \DEF{#6} 
     & #2 & \OR{#6} 
     & #3 & \OR{#6} 
     & #4 & \OR{#6} 
     & \BILD{17}{#6}{{\Large #5}}
  \end{tabular}}

\newcommand{\Gtrans}[6]%
  {\centering\begin{tabular}{ccc}
  #1 & \Garrow{5}{#3}{\to} & #2  \\[2mm]
  $\downarrow$ & & $\downarrow$ \\[2mm]
  #4 & \Garrow{5}{#6}{\To} & #5 
  \end{tabular}}

\begin{abstract}

  This paper illustrates how the \emph{diagram programming language}
  \textsc{DiaPlan} can be used to program visual systems.
  \textsc{DiaPlan} is a visual rule-based language that is founded
  on the computational model of graph transformation.  
  The language supports \emph{object-oriented programming}
  since its graphs are hierarchically structured.
  \emph{Typing} allows the shape of these graphs to be specified recursively
  in order to increase program security. 
  Thanks to its \emph{genericity}, \textsc{DiaPlan}
  allows to implement systems that represent and manipulate data
  in arbitrary diagram notations. 
  The environment for the language exploits the diagram editor generator
  \textsc{DiaGen} for providing genericity, 
  and for implementing its user interface and type checker.
\end{abstract}
\section{Introduction}\label{sec:intro}

Many data structures can be represented by graphs, 
and the theory of graph transformation~\cite{Rozenberg:97}
provides a profound computational model for rule-based programming with graphs.
As graphs are inherently visual, and have been used as an abstract model for
visual representations~\cite{Bardohl:98,Erwig-Meyer:95,Minas:00},
graph transformation could become widely accepted as a paradigm of 
\emph{rule-based visual programming}.

However, existing programming languages based on graph transformation, such
as {\sc Progres}~\cite{Schuerr-Winter-Zuendorf:99} or
$\textsc{Ludwig}_2$~\cite{Pfeiffer:95}, have not been as successful as one
could expect. We believe that this has two major reasons:
\begin{itemizeri}
\item The \emph{structuring} of graphs as nested graph objects, and thus
  \emph{object-oriented programming} are not supported.
\item It is not possible to \emph{customize} the ``standard'' graph
  notation to the visual notation of particular application domains.
\end{itemizeri}

However, some approaches to these problems exist for visual environments
which are not based on graphs: Object-oriented programming languages have
been developed in the visual programming community~\cite{VOOP:94}, notably
\emph{Prograph}~\cite{Prograph}. Furthermore, there are visual language
tools which allow to use domain-specific visual representations, e.g.,
\textsc{Calypso}, a tool for visually defining data structures
of programs~\cite{Wodtli-Cull:97}. However, we are not aware of any language
or tool that allows to \emph{visually specify} and \emph{generate} 
visual language environments which then use domain-specific visual
representations.

This paper is about \textsc{DiaPlan},  
a visual, rule-based programming language and environment for implementing visual languages, which offers just these features. 
The \textsc{DiaPlan} programming environment consists of
\begin{itemizeri}
\item a visual programming language which supports graph typing and
  structuring as well as object-oriented programming for specifying graph
  transformations. These rules specify the behavior of the generated
  visual language.
\item a tool for specifying how graphs are represented by domain-specific
  diagrams in the generated visual language and how the user can interact
  with these diagrams. This makes \textsc{DiaPlan} a \emph{generic}
  environment.
\end{itemizeri}

The paper is organized as follows: Section~\ref{s:hgratra} describes graph
transformation, the computational model of the language. Concepts for
programming are illustrated in Section~\ref{s:control}, and typing is
discussed in Section~\ref{s:typing}. A complete example of a
\textsc{DiaPlan} program for solving graph coloring problems is discussed
in Section~\ref{s:example}. Section~\ref{s:diagrams} recalls how graphs can
be visualized in arbitrary diagram notations. Genericity of the language is
based on this feature.
In Section~\ref{s:implementation} we describe how the language can be
implemented. We conclude in Section~\ref{s:conclusion} with some remarks on
related and future work.

Due to space limitations, our presentation can only be informal.
We explain the concepts by a running example
concerned with the visual representation of lists and 
the implementation of list operations.

\section{The Computational Model}\label{s:hgratra}

We introduce a rather general notion of graphs, and a rather simple way of applying rules to them.

\subsection{Graphs}\label{ss:hgraphs} 
\begin{figure*}[b]
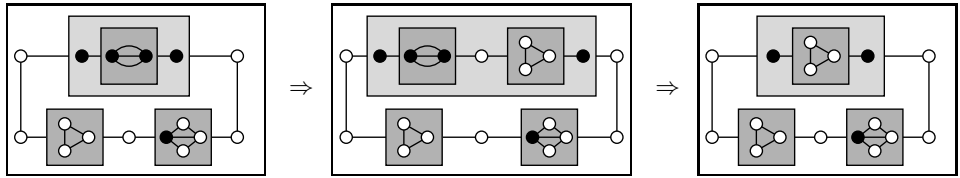
 %
\newcommand{\GE}
{\begin{framegraph}(38,25)(-2,-17.5)
  \node{h}{0}{0} \edge{h}{B} 
  \Qframe{B}{18}{12}{16}{0}{(17,11)(3.5,-5.5)
          \point{v0}{5}{0} \edge{v0}{Q} 
          \Iframe{Q}{8.5}{8.5}{12}{0}{\ItemII}
          \point{v1}{19}{0} \edge{v1}{Q}
          }  \edge{B}{t}
  \node{t}{32}{0}  
  \node{v4}{0}{-12} \edge{v4}{X} \edge{v4}{h}
  \Iframe{X}{8.5}{8.5}{8}{-12}{\ItemIII}
  \node{v5}{16}{-12} \edge{v5}{X} \edge{v5}{Y}
  \Iframe{Y}{8.5}{8.5}{24}{-12}{\ItemI}
  \node{v7}{32}{-12} \edge{v7}{Y} \edge{v7}{t}
\end{framegraph}
}
\newcommand{\HE}
{\begin{framegraph}(44,25)(-2,-17.5)
  \node{h}{0}{0} \edge{h}{B}
  \Qframe{B}{34}{12}{20}{0}{(33,11)(3.5,-5.5) 
          \point{v0}{5}{0} \edge{v0}{G1} 
          \Iframe{G1}{8.5}{8.5}{12}{0}{\ItemII}
          \node{v1}{20}{0} \edge{v1}{G2} \edge{v1}{G1} 
          \Iframe{G2}{8.5}{8.5}{28}{0}{\ItemIII} 
          \point{v2}{35}{0} \edge{v2}{G2}
          } \edge{B}{t} 
  \node{t}{40}{0}
  \node{v4}{0}{-12} \edge{v4}{X} \edge{v4}{h}
  \Iframe{X}{8.5}{8.5}{10}{-12}{\ItemIII}
  \node{v5}{20}{-12} \edge{v5}{X} \edge{v5}{Y}
  \Iframe{Y}{8.5}{8.5}{30}{-12}{\ItemI}
  \node{v7}{40}{-12} \edge{v7}{Y} \edge{v7}{t}
\end{framegraph}
}
\newcommand{\GD}
{\begin{graph}(48,23) \end{graph}
}
\newcommand{\HD}
{\begin{framegraph}(38,25)(-2,-17.5)
  \node{h}{0}{0} \edge{h}{B}
  \Qframe{B}{19}{12}{16}{0}{(18,11)(11,-5.5)
          \point{v0}{13}{0} \edge{v0}{G2} 
          \Iframe{G2}{8.5}{8.5}{20}{0}{\ItemIII} 
          \point{v2}{27}{0} \edge{v2}{G2}
          } \edge{B}{t} 
  \node{t}{32}{0}  
  \node{v4}{0}{-12} \edge{v4}{X} \edge{v4}{h}
  \Iframe{X}{8.5}{8.5}{8}{-12}{\ItemIII}
  \node{v5}{16}{-12} \edge{v5}{X} \edge{v5}{Y}
  \Iframe{Y}{8.5}{8.5}{24}{-12}{\ItemI}
  \node{v7}{32}{-12} \edge{v7}{Y} \edge{v7}{t}
\end{framegraph}
}
\STEPS{\GE}{\HE}{\HD}{25} %
\caption{Two transformation steps on list graphs}%
\label{f:enq-deq-steps}
\end{figure*}%

\emph{Graphs} represent relations between entities as \emph{edges} between
\emph{nodes}. Usually, the edges link two nodes of a graph, and represent
binary relations. We, however, allow edges that link \emph{any number} of
nodes, and distinguish different \emph{types} of edges by labeling them so
that different relations, of any arity, can be represented in a single
graph.  We also distinguish a sequence of nodes as the \emph{points} at which a graph may be connected to other graphs. 
Such graphs are known as \emph{pointed hypergraphs}~\cite{Drewes-Habel-Kreowski:97}.

Usually, graphs are \emph{flat}: Their nodes and edges are primitive; none
of them may contain a nested graph.  
We, however, distinguish a subset of the edges in a graph $H$ as \emph{frames}: Every frame $f$ \emph{contains} a nested subgraph $H_f$ that may contain frames again. 
These graphs are called \emph{hierarchical} in~\cite{Drewes-Hoffmann-Plump:00}.

\begin{exxx}[List Graphs]
\label{x:hlist}%
The three graphs in Figure~\ref{f:enq-deq-steps} show how we represent
lists as graphs: A \emph{list} frame (light gray boxes) is linked to the
start and end node of a chain of item frames (dark gray boxes); every
\emph{item} frame contains an \emph{item graph}.
 
Nodes are drawn as circles, and filled if they are points.  Edges are drawn
as boxes around their label, and are connected to their attachments by
lines that are ordered counter-clockwise, starting at noon.  The boxes for
binary edges with empty labels ``disappear'' so that they are drawn as
lines from their first to their second linked node.  Frames are boxes (like
ordinary edges), with their contents drawn inside; we distinguish list and
item frames by different shades of grey, and omit their labels. The graphs
in Figure~\ref{f:enq-deq-steps} contain two item frames, and a list frame
that contains one or two item frames.
\wzbw\end{exxx}

This representation \emph{forbids} edges across frame boundaries (which other notions of hierarchical graphs~\cite{Pratt:71,Engels-Schuerr:95} allow). 
Only then graphs can be transformed in a modular way. 
However, the correspondence between the links of a frame and the points of its contents may induce an \emph{indirect relation} between the contents and the context of that frame. 
The links of the list frames in Figure~\ref{f:enq-deq-steps} are related to the points of their contents in that way.

\subsection{Computation}\label{ss:rules}
In the graphs occurring in computation rules, a distinguished set $X$ of
\emph{variable names} is used to label \emph{variable edges}
(\emph{variables} for short). 
A \emph{substitution} $\sigma $ maps variables names $X_1, \ldots, X_n$ onto graphs $G_1, \ldots, G_n$.

The \emph{instantiation} $G\sigma$ of a graph $G$ by some substitution
$\sigma$ is obtained by identifiying the points of a fresh copy of $\sigma(X_i)$
with the corresponding attachments of every $X_i$-edge in $G$, and removing
the edge afterwards. 

A graph $C[\,]$ is called a \emph{context} if it contains a single variable (a \emph{hole}). 
We write $C[G]$ for the instantiation of the hole in $C[\,]$ by some graph $G$.

A \emph{graph transformation rule} (\emph{rule}, for short) $t: P \to R$
consists of a \emph{pattern} $P$ and a \emph{replacement} $R$, where $P$
and $R$ are graphs such that the following holds:
\begin{itemize}
\item Every variable name occurring in $R$ occurs in $P$ as well.
\item Every variable name occurs at most once in $P$.
\end{itemize}

A rule $t: P \to R$ \emph{transforms} some host graph $G$ into a modified
graph $H$, written $G \To_t H$, if there is a context $C[\,]$ and a
substitution $\sigma$ such that $G \iso C[P\sigma]$ and $H \iso
C[R\sigma]$. 
\footnote{We write $G \iso H$ if two graphs $G$ and $H$ are isomorphic, i.e.~equal up to the identity of their nodes and edges.}

A transformation step can be constructed by \emph{graph
  matching} (defined in~\cite{Plump-Habel:96} for flat graphs). 
The matching algorithm can be lifted to the hierarchical case along the lines of~\cite{Drewes-Hoffmann-Plump:00}.

\begin{exxx}[List Graph Transformation]%
\label{x:enremove} 
The rules in Figure~\ref{f:enq-deq-hrules} specify two operations on list graphs, which are used in the transformation steps shown in Figure~\ref{f:enq-deq-steps}.
(Variable names appear in italics.)
\begin{figure*}[bt]
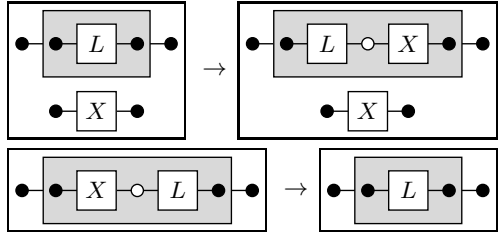
 %
\newcommand{\PE}
{\begin{framegraph}(26,20)(-2,-14)
  \point{h}{0}{0} \edge{h}{B} 
  \Qframe{B}{16}{10}{11}{0}{(16,10)(4,-5)
          \point{v0}{6}{0} \edge{v0}{Q} 
          \var{Q}{12}{0}{L}
          \point{v1}{18}{0} \edge{v1}{Q}
          }  \edge{B}{t}
  \point{t}{22}{0}  
  \point{v4}{5}{-10} \edge{v4}{X} 
  \framevar{X}{11}{-10}{X}
  \point{v5}{17}{-10} \edge{v5}{X} 
\end{framegraph}
}
\newcommand{\RE}
{\begin{framegraph}(38,20)(-2,-14)
  \point{h}{0}{0} \edge{h}{B}
  \Qframe{B}{28}{10}{17}{0}{(28,10)(4,-5) 
          \point{v0}{6}{0} \edge{v0}{G1} 
          \var{G1}{12}{0}{L}
          \node{v1}{18}{0} \edge{v1}{G2} \edge{v1}{G1} 
          \framevar{G2}{24}{0}{X} 
          \point{v2}{30}{0} \edge{v2}{G2}
          } 
  \edge{B}{t} 
  \point{t}{34}{0}

  \point{v4}{11}{-10} \edge{v4}{X} 
  \framevar{X}{17}{-10}{X}
  \point{v5}{23}{-10} \edge{v5}{X} 
\end{framegraph}
}
\newcommand{\PD}
{\begin{framegraph}(38,12)(-2,-6)
  \point{h}{0}{0} \edge{h}{B}
  \Qframe{B}{28}{10}{17}{0}{(28,10)(4,-5) 
          \point{v0}{6}{0} \edge{v0}{G1} 
          \framevar{G1}{12}{0}{X}
          \node{v1}{18}{0} \edge{v1}{G2} \edge{v1}{G1} 
          \var{G2}{24}{0}{L} 
          \point{v2}{30}{0} \edge{v2}{G2}
          } 
  \edge{B}{t} 
  \point{t}{34}{0}
\end{framegraph}
}
\newcommand{\RD}
{\begin{framegraph}(26,12)(-2,-6)
  \point{h}{0}{0} \edge{h}{B}
  \Qframe{B}{16}{10}{11}{0}{(16,10)(4,-5) 
          \point{v0}{6}{0} \edge{v0}{G1} 
          \var{G1}{12}{0}{L}
          \point{v1}{18}{0} \edge{v1}{G1} 
          } 
  \edge{B}{t} 
  \point{t}{22}{0}
\end{framegraph}
}
\RULE{\PE}{\RE}{20}
\hspace{1cm} %
\RULE{\PD}{\RD}{12}
\caption{Two rules specifying operations on list graphs}%
\label{f:enq-deq-hrules}
\end{figure*}%

The first step (using the rule on the left) \emph{enters} a copy of an item frame $X$ at the end of a list graph $L$, and the second step (using the rule on the right) \emph{removes} the first item frame from a list graph.
\wzbw\end{exxx} %

\emph{Graph transformation} with a set $\T$ of graph transformation rules
considers sequences of sequential transformation steps in arbitrary order,
and of arbitrary length.

By taking arbitrary graphs as input, and transforming them as long as
possible, graph transformation computes a \emph{function} on graphs. 
This function is \emph{partial} if certain graphs can be transformed infinitely, and \emph{nondeterministic} if a graph may be transformed in different ways.

\section{Programming}\label{s:control}

The computational model presented in Section~\ref{s:hgratra}
is extended by concepts for abstraction, control, and encapsulation.
This section gives only a brief account of these programming concepts. 
See~\cite{Hoffmann:00} for more motivation and details.

\subsection{Abstraction}

We consider certain labels as \emph{predicate names}.
An edge labeled by a predicate name is depicted as an oval. 
A \emph{predicate} named $p$ is defined by a set of rules wherein every pattern contains exactly one $p$-edge, and every replacement may contain other predicate edges. 
A predicate $p$ is \emph{applied} by applying one of its rules to a $p$-edge in the host graph.  
A predicate is \emph{evaluated} by first applying one of its rules, and evaluating all predicates that are called in its replacement, recursively.

The links of a predicate edge indicate the \emph{parameters} of a predicate. 
A parameter can be just a \emph{node}, but also an \emph{edge}. 
In particular, this edge can be a frame that contains a \emph{graph parameter}
(as in Example~\ref{x:removeproc} below), or a predicate edge that denotes a
\emph{predicate parameter} (as in Example~\ref{x:combinators} below). 

The rule set of a predicate may contain an \emph{otherwise} definition (starting with a ``$\others$'' symbol) that applies when no rule of the predicate is applicable.
In  Example~\ref{x:removeproc}, a ``--'' in the otherwise definition signals \emph{failure} of the predicate, and triggers backtracking, and in Example~\ref{x:combinators}, a ``+'' signals \emph{success} of the predicate, so that evaluation continues.
(A ``$\bot$'' can be used to raise an \emph{exception} that is either caught elsewhere, or leads to program abortion).

In any case, predicate edges are always removed during the transformation because
they are meta edges that are just introduced to control the program's
evaluation, and are not meant to occur in its result.

\begin{exxx}[A Predicate]\label{x:removeproc}
Figure~\ref{f:removeproc} shows the definition of a predicate \fg{remove}.
\begin{figure}[ht]
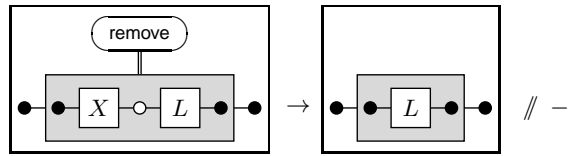
 %
\newcommand{\PD}
{\begin{framegraph}(38,21.5)(-2,-6.5)
  \button{D}{17}{11}{remove} \metaedge{D}{B}
  \point{h}{0}{0} \edge{h}{B}
  \Qframe{B}{28}{10}{17}{0}{(28,10)(4,-5) 
          \point{v0}{6}{0} \edge{v0}{G1} 
          \framevar{G1}{12}{0}{X}
          \node{v1}{18}{0} \edge{v1}{G2} \edge{v1}{G1} 
          \var{G2}{24}{0}{L} 
          \point{v2}{30}{0} \edge{v2}{G2}
          } 
  \edge{B}{t} 
  \point{t}{34}{0}
\end{framegraph}
}
\newcommand{\RD}
{\begin{framegraph}(26,21.5)(-2,-6.5)
  \point{h}{0}{0} \edge{h}{B}
  \Qframe{B}{16}{10}{11}{0}{(16,10)(4,-5) 
          \point{v0}{6}{0} \edge{v0}{G1} 
          \var{G1}{12}{0}{L}
          \point{v1}{18}{0} \edge{v1}{G1} 
          } 
  \edge{B}{t} 
  \point{t}{22}{0}
\end{framegraph}
}
\centering
\PROC{\fg{remove}}{\PD}{\RD}{-}{13}
\caption{The predicate \fg{remove}}%
\label{f:removeproc}
\end{figure}%
The predicate is parameterized by a frame (which should contain a list graph).
It is defined by a single rule (which appeared on the right in Figure~\ref{x:enremove}).
Its otherwise definition ``$\others -$'' leads to failure if it is applied to an empty list graph.

Note that the predicate \emph{updates} a list graph, as in imperative or object-oriented programming.
However, it would be possible to define a ``functional'' version of \fg{remove} that constructs a new list frame as a result, and leaves the input frame unchanged.
Such a predicate would need more space as subgraphs have to be copied (as in functional languages).
\wzbw\end{exxx} %

\subsection{Control}

Program evaluation is nondeterministic since predicates can be applied in arbitrary order, also concurrently.
We introduce \emph{conditional rules} by which an evaluation order for predicates can be specified. 
Such a rule has the form $t: P \guard A \to R$ where the graph $A$ is an \emph{application condition}  (or \emph{premise}).
It is applied to a graph $G$ as follows: 
If $G \iso C[P\sigma]$, the graph $C=G[(A \oplus R)\sigma]$ is constructed,
where $A \oplus R$ is the union of $A$ and $R$ that identifies only their
corresponding points. 
If all predicate edges of $A\sigma$ in $C$ can be evaluated, yielding a graph $H$, $t$ is applicable, and its application yields $H$; otherwise, the rule is not applicable, and $G$ is left unchanged.
(Note that the evaluation of $A\sigma$ may thus modify the host graph. 
          This effect is used in Example~\ref{x:combinators}.)

Predicates provide some simple control mechanisms:
Pattern matching and otherwise definitions allow for case distinction; applicability conditions specify an application order for predicates.

We also allow that predicates are parameterized by predicates.
This allows control to be specified by \emph{combinators} like in functional languages.
\begin{exxx}[Control Combinators]\label{x:combinators} 
  Figure~\ref{f:combinators} shows a predicate \fg{normalize} that
  applies to a predicate denoted by the variable $T$,
  evaluates $T$ as an application condition, and, if that succeeds, calls
  itself recursively.

  As $T$ shall bind to predicate calls with any number of parameters, we
  use the dot notation to indicate that $T$ links to a varying number of
  nodes.
  Where the $T$-edge is used as a predicate parameter, it is
  \emph{disguised} as an ordinary edge by drawing a box around the oval.
  This prevents it from evaluation as long as it is ``carried around'' (in the
  pattern and replacement graph of the rule).

\begin{figure}[th]
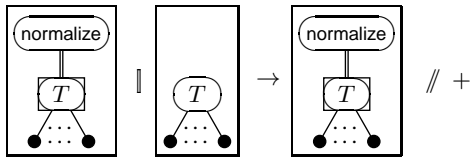
 %
\newcommand{\PN}
{\begin{framegraph}(16,22)(-2,0)
  \button{N}{6}{18}{normalize} \metaedge{N}{Q}
  \parbuttonvar{Q}{6}{9}{T}
  \edge{Q}{1} \edge{Q}{k}
  \hiddennode{d}{6}{4} \autonodetext{d}{\ldots}
  \point{1}{2}{2}
  \hiddennode{d2}{6}{2} \autonodetext{d2}{\ldots}
  \point{k}{10}{2}
\end{framegraph}
}
\newcommand{\AN}
{\begin{framegraph}(12,22)
  \buttonvar{Q}{6}{9}{T} \edge{Q}{1} \edge{Q}{k}
  \hiddennode{d}{6}{4} \autonodetext{d}{\ldots}[\opaquetextfalse]
  \point{1}{2}{2}
  \hiddennode{d2}{6}{2} \autonodetext{d2}{\ldots}[\opaquetextfalse]
  \point{k}{10}{2}
\end{framegraph}
}
\centering
\CONPROC{\fg{normalize}}{\PN}{\AN}{\PN}{+}{22}
\caption{The control combinator \fg{normalize}}%
\label{f:combinators}
\end{figure}%

\begin{figure*}[bt]
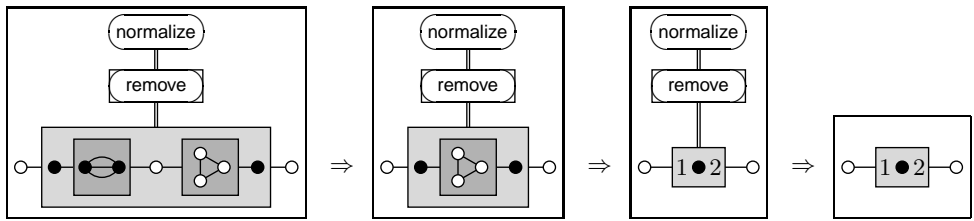
 %
\newcommand{\Neins}
{\begin{framegraph}(44,31)(-2,-7.5)
  \button{N}{20}{20}{normalize} \metaedge{D}{N}
  \parbutton{D}{20}{12}{remove} \metaedge{D}{B}
  \node{h}{0}{0} \edge{h}{B}
  \Qframe{B}{34}{12}{20}{0}{(33,11)(3.5,-5.5) 
          \point{v0}{5}{0} \edge{v0}{G1} 
          \Iframe{G1}{8.5}{8.5}{12}{0}{\ItemII}
          \node{v1}{20}{0} \edge{v1}{G2} \edge{v1}{G1} 
          \Iframe{G2}{8.5}{8.5}{28}{0}{\ItemIII} 
          \point{v2}{35}{0} \edge{v2}{G2}
          } 
  \edge{B}{t} 
  \node{t}{40}{0}
\end{framegraph}
}
\newcommand{\Nzwei}
{\begin{framegraph}(28,31)(-2,-7.5)
  \button{N}{12}{20}{normalize} \metaedge{D}{N}
  \parbutton{D}{12}{12}{remove} \metaedge{D}{B}
  \node{h}{0}{0} \edge{h}{B}
  \Qframe{B}{18}{12}{12}{0}{(17,11)(11.5,-5.5) 
          \point{v0}{13}{0} \edge{v0}{G2} 
          \Iframe{G2}{8.5}{8.5}{20}{0}{\ItemIII} 
          \point{v2}{27}{0} \edge{v2}{G2}
          } 
  \edge{B}{t} 
  \node{t}{24}{0}
\end{framegraph}
}
\newcommand{\Ndrei}
{\begin{framegraph}(20,31)(2,-7.5)
  \button{N}{12}{20}{normalize} \metaedge{D}{N}
  \parbutton{D}{12}{12}{remove} \metaedge{D}{B}
  \node{h}{4}{0} \edge{h}{B}
  \Qframe{B}{8}{6}{12}{0}{(8,6)(-4,-3) 
          \point{v0}{0}{0} \autonodetext{v0}[w]{$1$}\autonodetext{v0}[e]{$2$}  
          } 
  \edge{B}{t} 
  \node{t}{20}{0}
\end{framegraph}
}
\newcommand{\Nvier}
{\begin{framegraph}(20,15)(2,-7.5)
  \node{h}{4}{0} \edge{h}{B}
  \Qframe{B}{8}{6}{12}{0}{(8,6)(-4,-3) 
          \point{v0}{0}{0} \autonodetext{v0}[w]{$1$}\autonodetext{v0}[e]{$2$}  
          } 
  \edge{B}{t} 
  \node{t}{20}{0}
\end{framegraph}
}
\THREESTEPS{\Neins}{\Nzwei}{\Ndrei}{\Nvier}{15}
\caption{An evaluation of \fg{normalize}}%
\label{f:norm-eval}
\end{figure*}%
In Figure~\ref{f:norm-eval}, \fg{normalize} is applied to 
a disguised call of \fg{remove}.
Every application of \fg{normalize} removes one item frame 
by evaluating \fg{remove} as an application condition,
until the list frame contains no item frame, and \fg{remove} fails.
(The empty list graph is represented by a single node;
the numbers 1 and 2 attached to it shall indicate that this node is the first,
as well as the second point of the graph.)
This control combinator uses the side effect of evaluating the premise
reminding of the way how the \emph{cut} operator is used in \textsc{Prolog}. 

Figure~\ref{fig:not-seq} shows two other control combinators: \textsf{Seq}
specifies two predicates which have to be evaluated sequentially whereas
\textsf{not} actually does not modify anything; \textsf{not} fails if and
only if its argument can be evaluated, i.e., \textsf{not} specifies an
application condition which must not be satisfied. Please note that the
right-hand side of its rule is ``--''. i.e., failure, whereas its otherwise
definition specifies ``++'', i.e., success.  For an application of these
control combinators see Section~\ref{s:example}.
\begin{figure}[htb]
  \begin{center}
    \psfig{figure=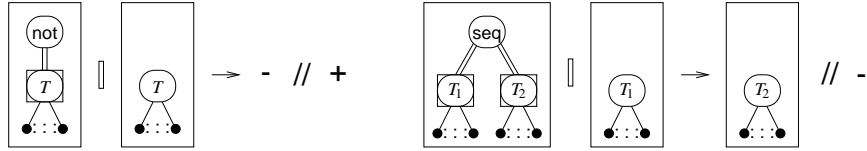,scale=0.6}
    \caption{The control combinators \textsf{not} and \textsf{seq}}
    \label{fig:not-seq}
  \end{center}
\end{figure}

Finally, Figure~\ref{fig:while} shows the two rules that define the control
combinator \textsf{while}: It has two parameters; the second one is
evaluated as long as the first one can be evaluated. The \textsf{while}
succeeds as soon as $T_1$ cannot be applied (any longer). The
\emph{otherwise} definition triggers backtracking if both rules fail, i.e.,
$T_1$ can be evaluated, but the evaluation of the replacement graph of the
lower rule fails. For an application of \textsf{not}, \textsf{seq}, and
\textsf{while} see Section~\ref{s:example}.
\begin{figure}[tb]
  \begin{center}
    \psfig{figure=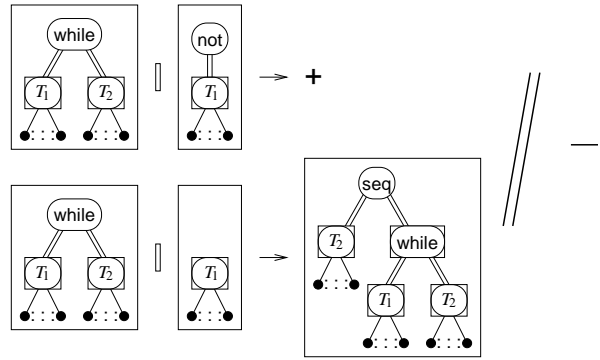,scale=0.6}
    \caption{The control combinator \textsf{while}}
    \label{fig:while}
  \end{center}
\end{figure}
\wzbw\end{exxx} %

\subsection{Encapsulation}

Programming--in--the--large relies on the encapsulation of features in modules 
so that only some of them are visible to the public, 
and the others are protected from illegal manipulation.

We consider frames as \emph{objects} that respond to certain \emph{messages} (which are predicate calls).
The types of frames are \emph{class names},
and a \emph{class definition} declares predicates as its \emph{methods}. 
Only the class name, and some designated methods are \emph{public}.
The graphs contained in frames, and their other methods, are \emph{private}.
This adheres to the principle of \emph{data abstraction}.

\begin{exxx}[The List Class] %
In Figure~\ref{f:class}
we encapsulate primitive operations on list graphs within a class.
\begin{figure}[ht]
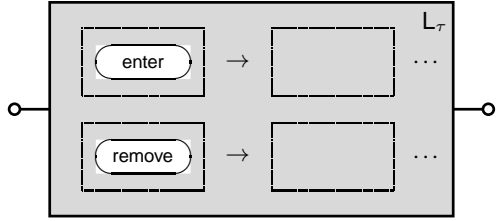
 %
\begin{center}
\begin{graph}(74,32)(-37,-16)
  \graphlinewidth*{2}
  \node{h}{-35}{0} \edge{h}{L}
  \Qframe{L}{60}{32}{0}{0}{(58,32)(-27,-16)
        \graphlinewidth*{0.5}
        \freetext(31,13){\makebox(0,6)[r]{\small \fg{L}$_\tau$}}
        \put(-23,2){\dashbox{0.2}(18,10){}} \freetext(0,7){$\to$}
        \put(5,2){\dashbox{0.2}(18,10){}} \freetext(28,7){\ldots}
        \button{E}{-14}{7}{enter}
        \put(-23,-12){\dashbox{0.2}(18,10){}} \freetext(0,-7){$\to$}
        \put(5,-12){\dashbox{0.2}(18,10){}} \freetext(28,-7){\ldots}
        \button{R}{-14}{-7}{remove}
          } 
  \node{t}{35}{0} \edge{t}{L}
\end{graph}
\end{center}
\caption{The list class}
\label{f:class}
\end{figure}%

In this small example, all methods are public.
However, the structure of list graphs is visible only inside for the methods \fg{remove} and \fg{enter} that belong to the class.
Other objects that contain some list frame $l$ can access its contents only by sending a message \fg{enter} or \fg{remove} to $l$.
\wzbw\end{exxx} %

\section{Typing}\label{s:typing}

Types classify the objects of a program, and establish rules for using
them. 
If these rules are checked, preferably \emph{statically}, before executing the program, this ensures that a program is consistent, and may also speed up
its execution.

\subsection{Graph Shapes}\label{ss:shapes}

We allow the \emph{shape} of graphs to be specified, similar as in \emph{Structured Gamma}~\cite{Fradet-LeMetayer:98}, by \emph{shape definitions} of the form
$$      G_T \mathrel{::=} G_1 \mid G_2 \mid \ldots \mid G_n     $$
where the graph $G_T$ consists of an edge labeled with a \emph{type name} $T$ and its linked nodes, 
and the $G_i$ are graphs that may contain type edges again.
Every variable name, frame type, and rule has a specified type,
and it is checked whether the substitutions of variables, the contents of frames, and the patterns and replacements of rules \emph{conform} to these types. 
Then all graphs in a program and the diagrams which are used as visual
representations are of a well-defined shape.

\begin{exxx}[Typing Shape of List and Item Graphs]%
\label{x:listtypes}
The shape definition in Figure~\ref{f:listtypes} specifies list and item graphs.
\begin{figure}[ht]
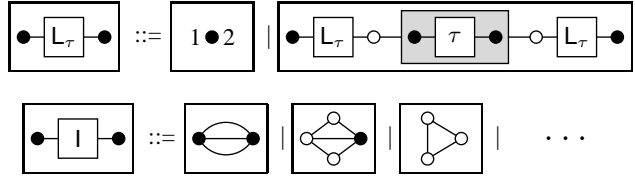
 %
\newcommand{\LQ}
 {\begin{framegraph}(16,10)(-2,13)
  \typ{Q}{6}{18}{L$_\tau$} \edge{Q}{b}  \edge{Q}{e} 
  \point{b}{0}{18} \point{e}{12}{18}
  \end{framegraph}
}
\TYPE{\LQ}
 {\begin{framegraph}(12,10)(-6,-5)
  \point{v}{0}{0} 
  \autonodetext{v}[w]{\small 1}\autonodetext{v}[e]{\small 2}
  \end{framegraph}
}{\begin{framegraph}(52,10)(-2,-5)
  \point{v0}{0}{0} \edge{v0}{Q1} 
  \typ{Q1}{6}{0}{L$_\tau$} 
  \node{v1}{12}{0} \edge{v1}{G2} \edge{v1}{Q1} 
  \Qframe{G2}{16}{8}{24}{0}{(16,8)(-2,-4)
        \point{b}{0}{0}
        \var{X}{6}{0}{\tau} \edge{X}{b} \edge{X}{e}
        \point{e}{12}{0}
        }
  \node{v2}{36}{0} \edge{v2}{Q3} \edge{v2}{G2}
  \typ{Q3}{42}{0}{L$_\tau$} 
  \point{v3}{48}{0} \edge{v3}{Q3}
  \end{framegraph}
}{10}
\\[3mm]
\TYPES{\begin{framegraph}(16,10)(-2,-5)
  \typ{I}{6}{0}{I} \edge{I}{1} \edge{I}{k}
  \point{1}{0}{0}
  \point{k}{12}{0}
  \end{framegraph}
}{\begin{framegraph}(12,10)(-6,-5)
  \point{0}{4}{0} \edge{0}{1} \bow{0}{1}{-0.3} \bow{0}{1}{+0.3}
  \point{1}{-4}{0}
  \end{framegraph}
}{\begin{framegraph}(12,10)(-6,-5)
  \point{0}{4}{0} \edge{0}{1} \edge{0}{2} \edge{0}{3}
  \node{1}{0}{-3} \edge{1}{3}
  \node{2}{0}{+3} \edge{2}{3}
  \node{3}{-4}{0}
\end{framegraph}
}{\begin{framegraph}(12,10)(-6,-5)
  \node{0}{3}{0} \edge{0}{1} \edge{0}{2} 
  \node{1}{-2}{-3} \edge{1}{2}
  \node{2}{-2}{+3} 
  \end{framegraph}
}{$\ldots$}{10}
\caption{The shape of list and item graphs}
\label{f:listtypes}
\end{figure}%
These graphs may be contained in list and item frames, respectively. The
type \fg{L}$_{\tau}$ of list graphs is \emph{polymorphic}.  The type
parameter $\tau$ can be instantiated with any shape. In our examples, the
variable $L$ binds list graphs of type \fg{L}$_{\tau}$, and the variable
$X$ binds graphs of the any type $\tau$, and is instantiated by the type
\fg{I} of item graphs in the transformations of Example~\ref{x:enremove}
and~\ref{x:combinators}.
\wzbw\end{exxx} %

The rules used for shape definitions are a well-studied special case of
\emph{context-free} graph transformation~\cite{Drewes-Habel-Kreowski:97}.
Type checking thus amounts to \emph{context-free graph parsing}, as
it is implemented in \textsc{DiaGen} (see Section~\ref{s:diagrams}).

Note that even if graph parsing may be expensive, it is done \emph{statically},  before executing the program, and will also reduce the search space for graph matching at runtime.

\subsection{Predicate Signatures}

The \emph{signature} of a predicate shall specify to which
kind of parameters it applies. As predicates are represented as graphs,
their signature can be specified by shape definitions for a designated type $\pi$ of predicates.

\begin{exxx}[Signature of List Predicates]%
\label{x:listsig}
In Figure~\ref{f:listsig}, we speci\-fy signatures for the predicates used
in our examples.
The predicate type $\pi$ has a varying number of parameter nodes
so that the rules of the shape definition 
have different left hand sides.
\begin{figure*}[tb]
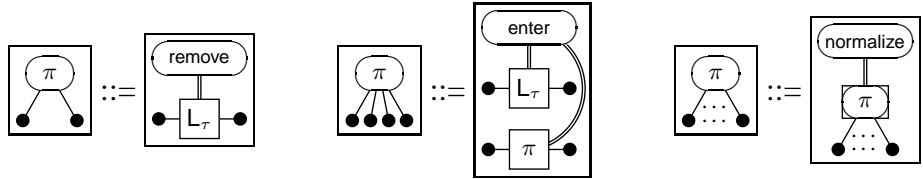
 %
\centering
\newcommand{\Nvariadic}
{\begin{framegraph}(16,21.5)(-2,0)
  \button{N}{6}{18}{normalize} \metaedge{N}{P}
  \parbuttonvar{P}{6}{9}{\pi}
  \edge{P}{1} \edge{P}{k}
  \hiddennode{d}{6}{4} \autonodetext{d}{\ldots}
  \point{1}{2}{2}
  \hiddennode{d2}{6}{2} \autonodetext{d2}{\ldots}
  \point{k}{10}{2}
\end{framegraph}
}
\newcommand{\Pvariadic}
{\begin{framegraph}(12,13)
  \buttonvar{P}{6}{9}{\pi} \edge{P}{1} \edge{P}{k}
  \hiddennode{d}{6}{4} \autonodetext{d}{\ldots}[\opaquetextfalse]
  \point{1}{2}{2}
  \hiddennode{d2}{6}{2} \autonodetext{d2}{\ldots}[\opaquetextfalse]
  \point{k}{10}{2}
\end{framegraph}
}
\newcommand{\Ntwosig}
{\begin{framegraph}(16,21.5)(-2,0)
  \button{N}{6}{18}{normalize} \metaedge{N}{P}
  \parbuttonvar{P}{6}{9}{\pi}
  \edge{P}{1} \edge{P}{k}
  \point{1}{2}{2}
  \point{k}{10}{2}
\end{framegraph}
}
\newcommand{\Ptwo}
{\begin{framegraph}(12,13)
  \buttonvar{P}{6}{9}{\pi} \edge{P}{1} \edge{P}{k}
  \point{1}{2}{2}
  \point{k}{10}{2}
\end{framegraph}
}
\newcommand{\Nfoursig}
{\begin{framegraph}(16,21.5)(-2,0)
  \button{N}{6}{18}{normalize} \metaedge{N}{P}
  \parbuttonvar{P}{6}{9}{\pi}
  \edge{P}{1} \edge{P}{2} \edge{P}{3} \edge{P}{k}
  \point{1}{2}{2}
  \point{2}{4.66}{2}
  \point{3}{7.36}{2}
  \point{k}{10}{2}
\end{framegraph}
}
\newcommand{\Pfour}
{\begin{framegraph}(12,13)
  \buttonvar{P}{6}{9}{\pi} \edge{P}{1} \edge{P}{2} \edge{P}{3} \edge{P}{k}
  \point{1}{2}{2}
  \point{2}{4.66}{2}
  \point{3}{7.36}{2}
  \point{k}{10}{2}
\end{framegraph}
}
\newcommand{\Dsig}
{\begin{framegraph}(16,16.5)(-2,5)
  \button{D}{6}{18}{remove} \metaedge{D}{Q}
  \typ{Q}{6}{9}{L$_\tau$} \edge{Q}{q1} \edge{Q}{q2}
  \point{q1}{0}{9}
  \point{q2}{12}{9}
\end{framegraph}
}
\newcommand{\Esig}
{\begin{framegraph}(17,25.5)(-2,-4)
  \button{D}{6}{18}{enter} \metaedge{D}{Q} \metabow{D}{X}{+0.45}
  \typ{Q}{6}{9}{L$_\tau$} \edge{Q}{q1} \edge{Q}{q2}
  \point{q1}{0}{9}
  \point{q2}{12}{9}
  \var{X}{6}{0}{\pi} \edge{X}{x1} \edge{X}{x2}
  \point{x1}{0}{0}
  \point{x2}{12}{0}
\end{framegraph}
}
\begin{tabular}{c} \Ptwo \end{tabular} {\large ::=}
\begin{tabular}{c} \Dsig \end{tabular}
\hfill
\begin{tabular}{c} \Pfour \end{tabular} {\large ::=}
\begin{tabular}{c} \Esig \end{tabular}
\hfill
\begin{tabular}{c} \Pvariadic \end{tabular} {\large ::=}
\begin{tabular}{c} \Nvariadic \end{tabular}

\caption{The signature of list predicates}
\label{f:listsig}
\end{figure*}%

All predicate calls occurring in the examples of this paper
can be derived with these rules 
(together with those of Figure~\ref{f:listtypes}).
The predicate variable $T$ used in example~\ref{x:combinators} is of type $\pi$.
\wzbw\end{exxx} %

Note how typing increases the security of programs:
Since the shape of all graphs in a program can be checked, every call of a predicate (like \fg{remove} or \fg{normalize}) can be type-checked at compile time.  
Since the input to the program, e.g.~the graphs in Figure~\ref{f:norm-eval}, can also be type-checked, a predicate will always be applied to graphs of its parameter type, and evaluation of the program may not go wrong with this respect.  

\section{Example: Graph Coloring}\label{s:example}

\begin{figure}[htb]
  \begin{center}
    \psfig{figure=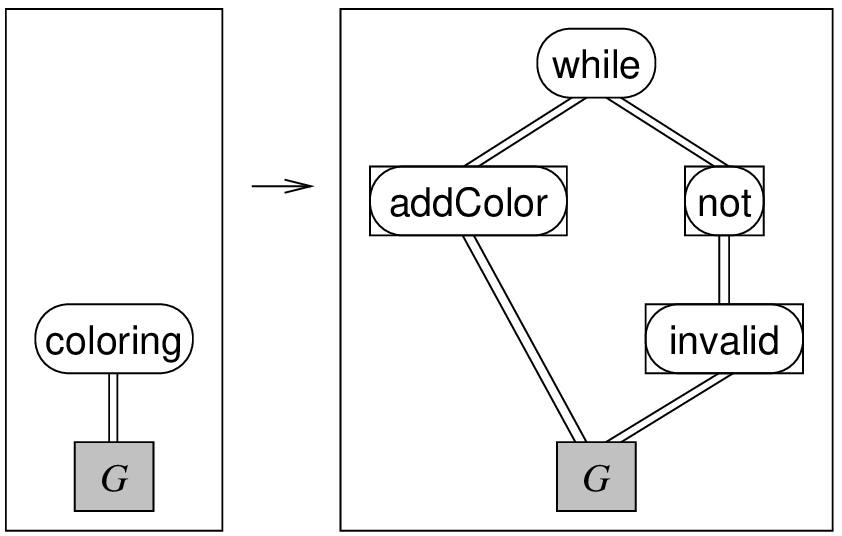,scale=0.6}
    \caption{The predicate \textsf{coloring}}
    \label{fig:coloring-pred}
  \end{center}
\end{figure}

This section shows that \textsc{DiaPlan} is a well-suited rule-based
language for inherently graphical problems. The following \textsc{DiaPlan}
program searches for a solution of the graph coloring problem for an
arbitrary graph which is passed as an argument to the predicate
\textsf{coloring} (cf.~Fig.~\ref{fig:coloring-pred}).
As specified by the \textsf{while}-combinator, the program alternately
evaluates predicate \textsf{addColor} and \textsf{not invalid}. The program
terminates as soon as \textsf{addColor} cannot be applied any longer
(success) or when no coloring exists (failure).

\begin{figure}[htb]
  \begin{center}
    \psfig{figure=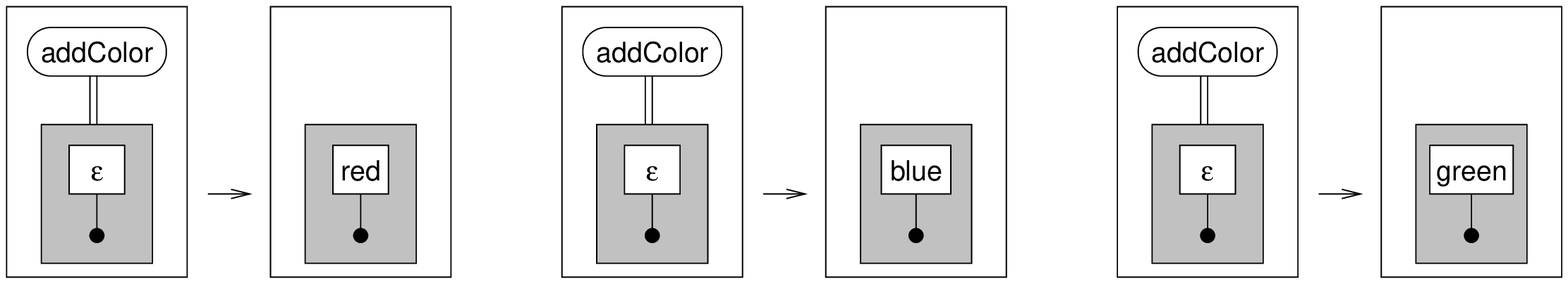,width=\textwidth}
    \caption{The predicate \textsf{addColor}}
    \label{fig:addColor-pred}
  \end{center}
\end{figure}
Predicate \textsf{addColor} (Fig.~\ref{fig:addColor-pred}) simply adds a
color edge, i.e., \emph{red}, \emph{blue}, or \emph{green} to some
previously non-colored node of the graph. Non-colored nodes are indicated by
$\varepsilon$-edges.
\begin{figure}[bh]
  \begin{center}
    \psfig{figure=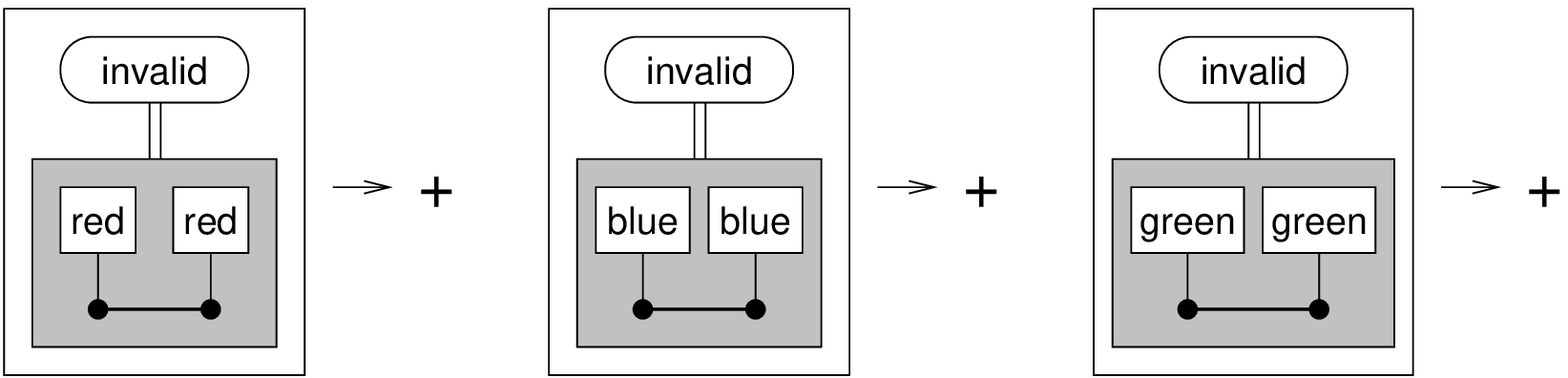,scale=0.6}
    \caption{The predicate \textsf{invalid}}
    \label{fig:invalid-pred}
  \end{center}
\end{figure}
After assigning a color to a node, predicate \textsf{invalid}
(Fig.~\ref{fig:invalid-pred}) checks whether this action was valid: The
\textsf{not invalid} predicate fails and triggers backtracking if previous
\textsf{addColor} evaluation cannot yield a consistent graph coloring.

\begin{figure}[htb]
  \begin{center}
    \psfig{figure=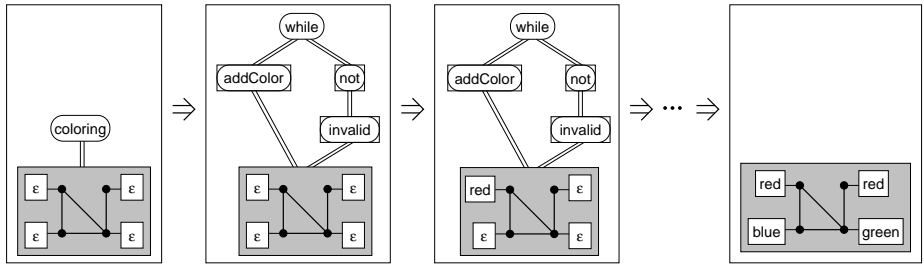,width=\textwidth}
    \caption{An evaluation of \textsf{coloring} which yields a graph coloring}
    \label{fig:color-ex}
  \end{center}
\end{figure}

Fig.~\ref{fig:color-ex} shows a sample transformation sequence
which terminates with a consistent coloring of a simple graph that
consists of four edges and four nodes which are initially non-colored.

\section{Diagrams}\label{s:diagrams}

Genericity, the other key feature of the language, is based on diagrams as
an external notation for graphs. This section briefly recalls how diagrams
can be represented by graphs (e.g., in diagram editors), and, vice versa,
how graphs can be visualized by diagrams.

\subsection{Representation and recognition of diagrams}\label{s:graph-rep}

\begin{figure}[b]
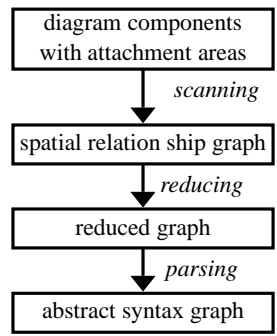
 %
\begin{center}
\unitlength=0.7mm\graphlinewidth*{3}\grapharrowwidth*{2}\grapharrowlength*{2}
\begin{graph}(50,60)(-20,-4)
  \rectnode{c}[50,12](0,52) 
        \autonodetext{c}{\small\begin{tabular}{@{}c@{}} diagram components\\ with attachment areas \end{tabular}}
  \diredge{c}{s} \bowtext{c}{s}{+0.7 }{\emph{scanning}}
  \rectnode{s}[50,8](0,32) 
        \autonodetext{s}{\small\begin{tabular}{@{}c@{}} spatial relation ship graph \end{tabular}}
  \diredge{s}{r} \bowtext{s}{r}{+0.7 }{\emph{reducing}}
  \rectnode{r}[50,8](0,16) 
        \autonodetext{r}{\small\begin{tabular}{@{}c@{}} reduced graph \end{tabular}}
  \diredge{r}{a} \bowtext{r}{a}{+0.7 }{\emph{parsing}}
  \rectnode{a}[50,8](0,0) 
        \autonodetext{a}{\small\begin{tabular}{@{}c@{}} abstract syntax graph \end{tabular}}
\end{graph}
\end{center}
\caption{Diagram representation and recognition in \textsc{DiaGen}}
\label{f:diagen}
\end{figure}

Andries \emph{et al.}~\cite{Andries-Engels-Rekers:98} have proposed a model
for representing diagrams as graphs that has been modified in
\textsc{DiaGen}, a tool for generating diagram editors which support
free-hand as well as syntax-directed
editing~\cite{Koeth-Minas:00,Minas:98,Minas:00}.  This model shall be
described here. Figure~\ref{f:diagen} shows the levels of diagram
representation and the recognition steps when analyzing a given diagram.

\emph{Scanning} creates a \emph{spatial relationship graph} (\emph{SRG} for
short) for capturing the lexical structure of a diagram. This step uses
edges for representing the \emph{elements} of a diagram language (like
circles, boxes, arrows, text); these \emph{component edges} are linked to
nodes representing the \emph{attachment areas} at which these elements can
be connected with the attachment areas of other elements, (like~the
\emph{border} and \emph{area} of circles and boxes, the \emph{source} and
\emph{target} ends of arrows). The connection of attachment areas is
explicitly represented by binary relationship edges. The type of a
relationship edge reflects the types of connected attachment areas and
their kind of connection (e.g., intersection). For instance, the
\emph{sources} and \emph{targets} of arrows may have intersections with the
\emph{borders} of circles or boxes leading to specific relationship edges
between the nodes of the corresponding element edges.

Syntactic analysis is performed in two steps: The SRG is first
\emph{reduced} to a more condensed graph (\emph{reduced graph}, RG for
short) which is then \emph{parsed} according to the syntax of the diagram
language. This two-step analysis transforms the spatial relationships into
logical relations of the \emph{abstract syntax graphs} of the diagrams.
(The specification of diagram syntax is discussed in
Section~\ref{s:typing}.) The separation into two independent steps allows
for a tractable process of specifying the syntax of the diagram language
and for efficient syntax analysis.

Syntax-directed editing is supported by such editors, too. \textsc{DiaGen}
comes with an abstract machine for the operational specification and
execution of graph transformation rules (cf.~Section~\ref{ss:rules}) which
are used to modify the SRG . This abstract machine already offers the full
functionality which is necessary for the implementation of an interpreter
for \textsc{DiaPlan} which is outlined in Section~\ref{s:implementation}.

This model is generic; a wide variety of diagram notations can be modeled,
e.g., finite automata and control flow diagrams~\cite{Minas-Viehstaedt:95},
Nassi-Shneiderman diagrams~\cite{Minas:97}, message sequence charts and
visual expression diagrams~\cite{Minas:98}, sequential function
charts~\cite{Koeth-Minas:00}, and ladder diagrams~\cite{Minas:00}. Actually
we are not aware of a diagram language that cannot be modeled that way. 

\subsection{Genericity}\label{s:genericity}
As this generic model uses graphs for modeling very different diagram
notations, we can also use diagrams for visualizing graphs. This capability
allows to tackle the problem that graphs are basically a visual data
structure, but using graphs for programming directly might be too abstract.
Instead, we can choose an arbitrary visual syntax for \emph{external}
representations even if the programming language represents visual data as
graphs internally.
The user interface of a program can so be customized for the visual
representations which are best suited in its application domain. This makes
it possible to use the programming language of this paper which is based on
graph transformations as a \emph{generic visual programming language}.  By
representing very different diagram notations by graphs and operating on
these graphs, many different flavors of visual (programming) languages can
be described and implemented.  Obvious examples are \emph{Pictorial
  Janus}~\cite{Kahn-Saraswat:90} (whose agents with ports directly
correspond to our notion of typed edges) or
\emph{KidSim}~\cite{Smith-Cypher-Spohrer:94}.

\section{Implementation}\label{s:implementation}

\begin{figure}[bt] %
  \centerline{\psfig{figure=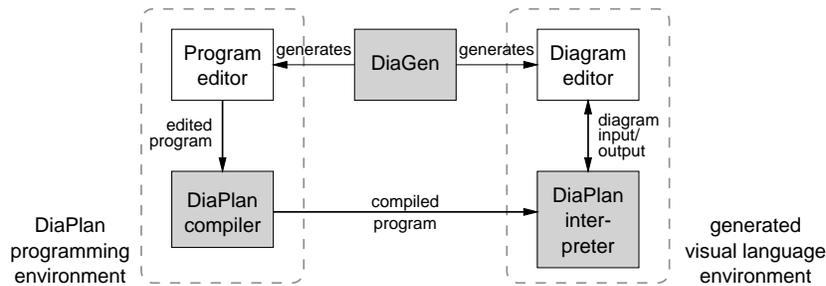,width=0.9\textwidth}}
\caption{System architecture}
\label{f:architecture}
\end{figure}

The programming language outlined in this paper is in a rather concrete phase of its design.
Its implementation is only at a very early stage.
Here we just outline the architecture of the implementation. 
(See Figure~\ref{f:architecture} for a diagram of the system structure.)

\begin{itemize}
\item The \emph{interpreter} executes programs of the language by reading
  the input graph, transforming the graph according to the program, and
  re-displaying the modified graph, in a loop, steered by user interaction.
  As an implementation of this interpreter, \textsc{DiaPlan} will use the
  abstract machine for graph transformations which is part of
  \textsc{DiaGen}.
\item The \emph{compiler} reads programs and transforms them into an
  internal form that can be easily and efficiently executed by the
  interpreter.  And, very important, the compiler checks whether the
  program violates any lexical, syntactical or contextual rule of the
  language.  Among the contextual rules, typing (as discussed in
  Section~\ref{s:typing}) plays a prominent role: The type checker shall be
  implemented with the graph parser built into \textsc{DiaGen}~\cite{Minas:97}. 
\item The interpreter shall have an interactive visual editor by which the
  input data is created. This editor shall be generated from a
  specification in \textsc{DiaGen} in order to support customizable diagram
  notation for the (abstract syntax) graphs that are produced by the editor
  in order to be transformed internally.
\item The programs will also be constructed with an interactive editor for
  the visual syntax of the programming language. Again, this editor shall
  be generated from a \textsc{DiaGen} specification of graphs, rules,
  predicates, types, and classes.
\end{itemize}

Altogether, only the compiler has to be implemented anew, while we rely
entirely on the operational graph transformation machine of {\sc DiaGen}
and the capability of \textsc{DiaGen} for generating the user interfaces of
\textsc{DiaPlan}. \textsc{DiaGen} needs some extensions to meet the needs
of this application:
\begin{itemize}
\item Other ways of specifying diagram languages, like normalizing
  constructor rules, have to be investigated.
\item A visual user interface has to be provided for \textsc{DiaGen} itself.
  It comes to no surprise that \textsc{DiaGen} shall be used for this purpose.
\end{itemize}

The implementation of the compiler is a challenging task. Even
if we have convinced the reader that all concepts promised for the language
are implementable, neither does this mean that it can be done
\emph{efficiently}, nor that this will result in \emph{efficient systems}.
For instance, the matching of a rule explained in section~\ref{ss:rules}
requires to check \emph{subgraph isomorphism}, which is NP-hard in general.
We hope that we will come close the performance of logical and
functional languages' implementations, at least after restricting the shapes of graphs in a suitable way. 
However, we cannot draw very much from the experience with implementing textual (functional or logical) languages for this aspect of the implementation.
\section{Conclusion}\label{s:conclusion}

In this paper we have presented a new programming language based on graph
transformation that is rule-oriented, object-oriented, and supports
structured graphs.  In particular, we have discussed typing of the
language, and its genericity with respect to diagram notations. Genericity
allows to represent graphs by specific notations of the application domain in the user interface. This feature makes the language and its specified environment well-suited for simulations and animations.

\emph{Structured graphs} have already been proposed in the context of graph
transformation~\cite{Pratt:71,Engels-Schuerr:95}, in graphical data base
languages~\cite{Poulovassilis-Levene:94}, and in system modelling
languages~\cite{Tapken:99}. 
Graph \emph{shapes} exist in \emph{Structured Gamma}~\cite{Fradet-LeMetayer:98} (for unstructured graphs).

However, we are not aware of any other language or language proposal that
features structured graphs, transformations, and classes together with genericity and typing.

The precise definitions of the concepts presented in this paper has been
started in~\cite{Drewes-Hoffmann-Plump:00}, and needs to be continued. Some
more concepts, like \emph{concurrency} and \emph{distribution}, have still
to be considered.  As these concepts have been studied
by~\cite{Taentzer:96} and~\cite{Taentzer-Schuerr:95} in a similar setting,
there is some hope that these results can be extended to our language.

Last but not least, a compiler (with type checker) has still to be implemented.


\begin{thebibliography}{10}

\bibitem{Andries-Engels-Rekers:98}
M.~Andries, G.~Engels, and J.~Rekers.
\newblock How to represent a visual specification.
\newblock In K.~Marriott and B.~Meyer, editors, {\em Visual Language Theory},
  chapter~8, pages 245--260. Springer, New York, 1998.

\bibitem{Bardohl:98}
R.~Bardohl.
\newblock \textsc{GenGEd}: A generic graphical editor for visual languages
  based on algebraic graph grammars.
\newblock In {\em Proc. 1998 IEEE Symp.{} on Visual Languages (VL'98), Halifax,
  Canada}, pages 48--55, 1998.

\bibitem{VOOP:94}
M.~M. Burnett, A.~Goldberg, and T.~G. Lewis, editors.
\newblock {\em Visual Object-Oriented Programming}.
\newblock Manning, 1994.

\bibitem{Corradini-Montanari:95}
A.~Corradini and U.~Montanari, editors.
\newblock {\em Proc.\ Joint {\sc CompuGraph}/{\sc Semagraph} Workshop on Graph
  Rewriting and Computation}, number~2 in Electronic Notes in Theoretical
  Computer Science, {\tt http://www.elsevier.nl/locate/entcs}, 1995. Elsevier.

\bibitem{Prograph}
P.~T. Cox, F.~R. Giles, and T.~Pietrzykowski.
\newblock Prograph.
\newblock In Burnett et~al. \cite{VOOP:94}, chapter~3, pages 45--66.

\bibitem{Drewes-Habel-Kreowski:97}
F.~Drewes, A.~Habel, and H.-J. Kreowski.
\newblock Hyperedge replacement graph grammars.
\newblock In Rozenberg \cite{Rozenberg:97}, chapter~2, pages 95--162.

\bibitem{Drewes-Hoffmann-Plump:00}
F.~Drewes, B.~Hoffmann, and D.~Plump.
\newblock Hierarchical graph transformation.
\newblock In {\em Foundations of Software Science and Computation Structures
  (FOSSACS 2000)}, LNCS, 2000.
\newblock To appear.

\bibitem{Engels-Schuerr:95}
G.~Engels and A.~Sch\"urr.
\newblock Encapsulated hierachical graphs, graph types, and meta types.
\newblock In Corradini and Montanari \cite{Corradini-Montanari:95}.

\bibitem{Erwig-Meyer:95}
M.~Erwig and B.~Meyer.
\newblock Heterogeneous visual languages -- {I}ntegrating visual and textual
  programming.
\newblock In {\em Proc.\ 11th IEEE Symp. on Visual Languages (VL'95),
  Darmstadt, Germany}, pages 318--325, 1995.

\bibitem{Fradet-LeMetayer:98}
P.~Fradet and D.~L. M{\'e}tayer.
\newblock Structured {G}gamma.
\newblock {\em Science of Computer Programming}, 31(2/3):263--289, 1998.

\bibitem{Hoffmann:00}
B.~Hoffmann.
\newblock From graph transformation to rule-based programming with diagrams.
\newblock In {\em Proc. Int. Workshop on Applications of Graph Transformations
  with Industrial Relevance ({\sc Agtive}'99)}, volume 1779 of {\em LNCS},
  pages 165--180.

\bibitem{Kahn-Saraswat:90}
K.~M. Kahn and V.~Saraswat.
\newblock Complete visualizations of concurrent programs and their executions.
\newblock In {\em Proc. IEEE Workshop on Visual Languages (VL'90)}, pages
  7--15, 1990.

\bibitem{Koeth-Minas:00}
O.~K\"{o}th and M.~Minas.
\newblock Generating diagram editors providing free-hand editing as well as
  syntax-directed editing.
\newblock In {\em Proc. International Workshop on Graph Transformation ({\sc
  GraTra} 2000), Berlin}, March 2000.

\bibitem{Minas:00}
M.~Minas.
\newblock Creating semantic representations of diagrams.
\newblock In {\em Proc. Int. Workshop on Applications of Graph Transformations
  with Industrial Relevance ({\sc Agtive}'99)}, volume 1779 of {\em LNCS},
  pages 209--224.

\bibitem{Minas:98}
M.~Minas.
\newblock Hypergraphs as a uniform diagram representation model.
\newblock In {\em Proc. 6th Int. Workshop on Theory and Application of Graph
  Transformation (TAGT'98)}, volume 1764 of {\em LNCS}, pages 281--295.

\bibitem{Minas:97}
M.~Minas.
\newblock Diagram editing with hypergraph parser support.
\newblock In {\em Proc.\ 13th IEEE Symp. on Visual Languages (VL'97), Capri,
  Italy}, pages 230--237, 1997.

\bibitem{Minas-Viehstaedt:95}
M.~Minas and G.~Viehstaedt.
\newblock {DiaGen}: A generator for diagram editors providing direct
  manipulation and execution of diagrams.
\newblock In {\em Proc.\ 11th IEEE Symp. on Visual Languages (VL'95),
  Darmstadt, Germany}, pages 203--210, 1995.

\bibitem{Pfeiffer:95}
J.~J. {Pfeiffer, Jr.}
\newblock $\textsc{Ludwig}_2$: Decoupling program representations from
  processing models.
\newblock In {\em Proc.\ 11th IEEE Symp. on Visual Languages (VL'95),
  Darmstadt, Germany}, pages 133--139, 1995.

\bibitem{Plump-Habel:96}
D.~Plump and A.~Habel.
\newblock Graph unification and matching.
\newblock In J.~E. Cuny, H.~Ehrig, G.~Engels, and G.~Rozenberg, editors, {\em
  Proc.\ Graph Grammars and Their Application to Computer Science}, number 1073
  in Lecture Notes in Computer Science, pages 75--89. Springer, 1996.

\bibitem{Poulovassilis-Levene:94}
A.~Poulovassilis and M.~Levene.
\newblock A nested-graph model for the representation and manipulation of
  complex objects.
\newblock {\em ACM Transactions on Information Systems}, 12(1):35--68, 1994.

\bibitem{Pratt:71}
T.~W. Pratt.
\newblock Pair grammars, graph languages and string-to-graph translations.
\newblock {\em Journal of Computer and System Sciences}, 5:560--595, 1971.

\bibitem{Rozenberg:97}
G.~Rozenberg, editor.
\newblock {\em Handbook of Graph Grammars and Computing by Graph
  Transformation, Vol.~{I}: Foundations}.
\newblock World Scientific, Singapore, 1997.

\bibitem{Schuerr-Winter-Zuendorf:99}
A.~Sch\"urr, A.~Winter, and A.~Z{\"u}ndorf.
\newblock The {\sc progres} approach: Language and environment.
\newblock In Rozenberg \cite{Rozenberg:97}, chapter~13, pages 487--550.

\bibitem{Smith-Cypher-Spohrer:94}
D.~C. Smith, A.~Cypher, and J.~Spohrer.
\newblock Kid{S}im: Programming agents without a programming language.
\newblock {\em Communications of the ACM}, 37(7):54--67, 1994.

\bibitem{Taentzer:96}
G.~Taentzer.
\newblock {\em Parallel and Distributed Graph Transformation: Formal
  Description and Application to Communication-Based Systems}.
\newblock Dissertation, TU Berlin, 1996.
\newblock Shaker Verlag.

\bibitem{Taentzer-Schuerr:95}
G.~Taentzer and A.~Sch{\"u}rr.
\newblock {DIEGO}, another step towards a module concept for graph
  transformation systems.
\newblock In Corradini and Montanari \cite{Corradini-Montanari:95}.

\bibitem{Tapken:99}
J.~Tapken.
\newblock Implementing hierarchical graph structures.
\newblock In J.-P. Finance, editor, {\em Proc. Formal Aspects of Software
  Engineering (FASE'99)}, number 1577 in Lecture Notes in Computer Science.
  Springer, 1999.

\bibitem{Wodtli-Cull:97}
R.~Wodtli and P.~Cull.
\newblock \textsc{Calypso}: A visual language for data structures programming.
\newblock In {\em Proc. 1997 IEEE Symp.{} on Visual Languages (VL'97), Capri,
  Italy}, pages 166--167, 1997.

\end{thebibliography}
\end{document}